\documentclass{JINST}
\let\ifpdf\relax
\usepackage{url}
\usepackage{tabularx}
\newcommand{\figref}[1]{\figurename~\ref{#1}}
\newcommand{\secref}[1]{Section~\ref{#1}}

\newcommand{\equref}[1]{Eq.~\ref{#1}}
\newcommand{\tabref}[1]{Table~\ref{#1}}

\usepackage{booktabs}
\usepackage{xspace}

\newcommand{\dqdxz}{\ensuremath{\Delta Q_0/\Delta s_0}\xspace}
\newcommand{\dqdxi}{\ensuremath{\Delta Q_i/\Delta s_i}\xspace}
\newcommand{\dqdxo}{\ensuremath{\Delta Q_1/\Delta s_1}\xspace}

\title{Long-term operation of a double phase LAr LEM Time Projection Chamber
with a simplified anode and extraction-grid design}

\author{C.~Cantini$^a$, L.~Epprecht$^a$,
  A.~Gendotti$^a$, S.~Horikawa$^a$,
  S.~Murphy$^a$, G.~Natterer$^a$,  L.~Periale$^a$, F.~Resnati$^a$,
  A.~Rubbia$^a$\thanks{Corresponding
    author.}, F.~Sergiampietri$^{a,b}$, T.~Viant$^a$ and S.~Wu$^a$~\\
  \llap{$^a$}ETH Zurich, Institute for Particle Physics,\\
  CH-8093 Z\"{u}rich, Switzerland\\
  \llap{$^b$}INFN-Sezione di Pisa\\
   56127 Pisa, Italy\\
  E-mail: \email{Andre.Rubbia@cern.ch}}

\abstract{We report on the successful operation of a double phase
  Liquid Argon Large Electron Multiplier Time Projection Chamber (LAr
  LEM-TPC) equipped with two dimensional projective anodes with
  dimensions 10$\times$10 cm$^2$, and with a maximum drift length of
  21~cm.  The anodes were manufactured for the first time from a
  single multilayer printed circuit board (PCB).  Various layouts of
  the readout views have been tested and optimised.  In addition, the
  ionisation charge was efficiently extracted from the liquid to the
  gas phase with a single grid instead of two previously. 
  We studied the response and the gain of the detector to cosmic muon tracks. 
  To study long-term stability over
  several weeks, we continuously operated the chamber at fixed
  electric field settings.  We reproducibly observe that after an initial decrease
  with a characteristic time of $\tau\approx 1.6$ days, the observed
  gain is stable. In 46~days of operation, a total of
  14.6~million triggers have been collected at a stable effective gain of $G_\infty\sim 15$ corresponding to a
  signal-to-noise ratio $(S/N)\gtrsim 60$ for minimum ionising
  tracks. During the full period, eight discharges across the LEM were observed.
  A maximum effective gain of 90 was also observed, corresponding to a signal-to-noise ratio
  $(S/N)\gtrsim 400$ for minimum ionising tracks, 
  or $S/N\approx10$ for an energy deposition of 15~keV on a single
   readout channel.  
  }


\begin{document}

\section{Introduction}
\label{sec:introduction}

The liquid Argon time projection chamber (LAr-TPC)~\cite{Amerio:2004} is a
charge imaging detector which allows to reconstruct tracks in
three dimensions as well as the locally deposited energy. In this context, 
the Giant Liquid Argon Charge Imaging ExpeRiment (GLACIER) is a
concept proposed for a future observatory for neutrino physics and
nucleon decay searches
~\cite{Rubbia:2004tz,Rubbia:2009md,Badertscher:2010sy}, which
could be scalable up to gigantic masses. This design is contemplated
for 20~kton and 50~kton detectors
in the LBNO Expression of Interest submitted to CERN~\cite{Stahl:2012exa}.

The key and innovative feature of the GLACIER design is the double
phase LEM-TPC operation mode with adjustable gain and 2D projective
readout~\cite{Badertscher:2013wm,Badertscher:2011sy,Badertscher:2010fi,Badertscher:2009av,Badertscher:2008rf}.
The ionisation charge is extracted to the Argon gas phase where it is
amplified by a Large Electron Multiplier (LEM) which triggers Townsend
multiplication in the high electric field regions in the LEM
holes~\cite{Badertscher:2008rf,Bondar:2008yw}.  The charge is
collected and recorded on a two-dimensional and segmented anode. This
principle has two main advantages: 1. the gain in the LEM is
adjustable, i.e. the signal amplitude can be optimised for $>99\%$ hit
reconstruction efficiency, and 2. the
signals collected on the two readout views are unipolar and symmetric
which facilitates the event reconstruction.  In addition to the
ionisation charge, scintillation light is emitted by excited Argon
diatomic molecules (excimers) (see e.g. Ref.~\cite{Boccone:2009kk}). The detection of the scintillation
light is fast, thus providing the event time reference $T_0$.  For
recent reviews on cryogenic detectors with electron avalanching see
e.g. Ref.~\cite{Buzulutskov:2011de,Chepel:2012sj}.

We have described the successful operation of this kind of detector in previous
publications (see for instance Ref.~\cite{Badertscher:2013wm,Badertscher:2011sy}). 
In this paper, we report on a simplified
design of  two
  dimensional projective readout anodes,
  manufactured
  from a single multilayer printed circuit board (PCB),
  leading to a reduction of the electrical capacitance of its electrodes.
  To further simplify the design,
we also decided to extract the drifting
electrons from the liquid to the gas by means of a single grid placed
just below the liquid surface. 
  These developments have been carried out in
view of the extrapolation of this technique to large surfaces
(typically one square meter), as needed for the GLACIER design.
The new readouts were manufactured and tested on a prototype chamber located
at CERN. The response of the chamber in terms of effective
gain, signal-to-noise ratio and energy resolution was
checked for different settings of the electric fields across each
stage. Furthermore, a stable operation of the TPC allowed us to
retrieve a data set of approximately 14.6~million events. The large
statistics enabled us to study the gain stability over longer periods.

\section{The experimental setup}
\begin{figure*}[h!]
    \centering
    \includegraphics[width=0.8\textwidth]{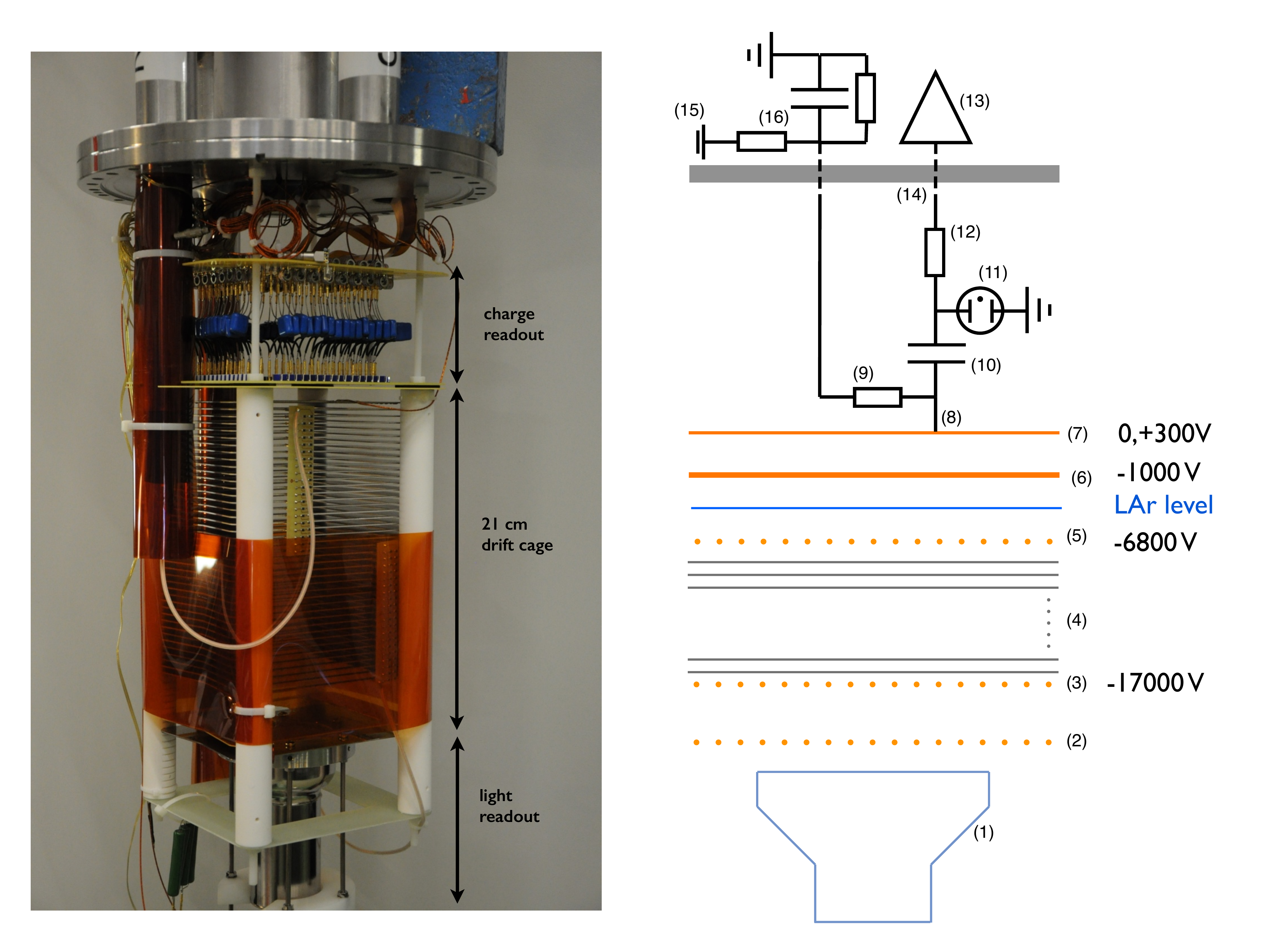}
    \caption{Left: Picture of the ETHZ 3 liter double phase LEM-TPC.  Right: schematic
      representation of the detector showing the PMT (1) and its
      protection grid (2), the cathode (3), the field shaping
      electrodes (4), the single extraction grid (5), the LEM (6), the
      two views low capacitance anode (7), the connection to the
      strips (8), the 500 M$\Omega$ high voltage resistor (9), 270 pF
      high voltage decoupling capacitor (10), the surge arrester (11),
      the current limiting 33 $\Omega$ resistor (12), the preamplifier
      (13), the flange and the electrical feedthroughs (14), the high
      voltage power supply (15), the low pass filter and the resistor
      divider (16). Elements from (8) to (13) are replicated for each
      readout channel. The voltages indicated on the right are those
      applied across each stage for the LEM operated at 33 kV/cm.}
    \label{fig:setup}
  \end{figure*}

The so-called ``3 liter'' setup is a double phase LAr LEM
TPC~\cite{Badertscher:2008rf,filippo_thesis, devis_thesis} consisting
of a 21~cm long drift volume and a 10$\times$10 cm$^2$ area. 
Even if of modest size, it is very
useful for testing new ideas with a rapid turn-around. The setup has
been developed by the ETHZ group during the last four years and
various charge readout devices were reliably operated with it, as
reported elsewhere~\cite{Badertscher:2011sy,Badertscher:2010fi}. A
picture of the chamber and the schematics of the setup are presented
in \figref{fig:setup}.  The ionisation electrons drift vertically
towards the liquid surface by means of a homogenous electric field (of
strength typ. 500~V/cm) provided by field shapers spaced out every 5~mm. 
A larger electric field of about 2 kV/cm is applied at the
vicinity of the liquid Argon surface to efficiently extract the
drifting charges to the vapour phase. The extraction field is confined
in a 10~mm region between the Large Electron Multiplier (LEM) and a
stainless steel etched mesh placed in the liquid (see
\secref{sec:singe_grid}).  The amplification and readout of the
charge is then performed by the LEM and anode layers.  The 1~mm thick
LEM used in this setup has about 16'000 holes of 500~$\mu$m spaced at
a distance of 800~$\mu$m from each other with a dielectric rim of
50~$\mu$m, and an active area which matches that of the readout. An
electric field of at least 30~kV/cm is applied across the LEM in order
to obtain charge amplification. The electrons are collected on the
anode placed 2~mm above the LEM.  The digitisation and data
acquisition is performed by the specially developed CAEN SY2791
readout system (see Ref.~\cite{Badertscher:2013wm}).  The detector is
also equipped with a Hamamatsu R11065 photomultiplier tube (PMT)
placed below the cathode grid and made sensitive to the VUV
scintillation light with the wavelength shifter tetraphenylbutadiene
(TPB) \cite{Boccone:2009kk}. The PMT provides the event $T_0$ and
trigger by detecting the prompt photons from the liquid Argon
scintillation.

In order to keep the 3 liter chamber at thermal equilibrium, it is fully
immersed in an open bath filled with liquid argon. The bath evaporates
to open air, the pressure and temperature are rather stable until
the level of the liquid argon bath goes below the top flange of the 3 liter
chamber. At this point, the heat input increases and the pressure
inside the chamber rises. The pressure inside the detector increases
until the operator refills the liquid argon bath. This is done periodically, on a daily basis.
In addition, the changes of atmospheric pressure are also reflected
on the temperature and pressure of the setup since the bath is open.
These pressure changes will affect the amplification in the LEM as will be shown
in Section~\ref{sec:stabgain}, however this effect on the gain can be corrected for
during offline data-analysis. 
To avoid such effects in future detectors,
it is straightforward to regulate the pressure avoiding open bath, as was for example 
the case in our 200~liter setup~\cite{Badertscher:2013wm}, where the pressure was stabilised
to $\pm 1$~mbar.

Before filling the inner vessel with (pure) liquid Argon, the volume is evacuated to
residual pressures below 5$\times10^{-6}$ mbar to favor the
outgassing of the materials. During the filling the liquid Argon is
passed through an activated copper and zeolite powder cartridge which traps the oxygen
and moisture impurities from the liquid. The filling is performed slowly, until
the level of the liquid is precisely adjusted inbetween the grid and the LEM. 
To maintain the purity for long periods,
the liquid Argon is evaporated and pushed by a bellows pump through a
commercial SAES getter\footnote{\url{www.saesgetters.com}}. The gas
Argon is then condensed through a serpentine immersed in a liquid
Argon open bath, and again reintroduced as liquid inside the
detector. The liquid Argon purity is monitored throughout the data
taking period by measuring the lifetime of the drifting electrons.

\subsection{New 2D views, low-capacitance anodes}\label{sec:anode_comp}
The anode consists of two orthogonal sets of strips (views) that
provide reconstruction of the $x$ and $y$ coordinates of the ionising
event. It is designed in such a way that the amplified charge is
(equally) shared and collected on both views. This feature simplifies the
extraction of the signal waveforms and eases the event reconstruction
process. 

In our previous designs~\cite{Badertscher:2011sy,filippo_thesis}, 
symmetric charge sharing was achieved by producing
anodes which consisted of two orthogonal sets of copper strips
interleaved with a 50~$\mu$m thick Kapton insulator. Those anodes,
referred to hereafter as Kapton foil anodes, had a measured
capacitance per unit length as high as $\sim$600 pF/m due the
proximity of the copper strips. 
As described in Ref.~\cite{Badertscher:2013wm}, the measured RMS value
of the intrinsic equivalent input noise charge (ENC) for a capacity
$C$ at the input of our preamplifiers is $470 \pm 30$ $e^-$ for
$C=10$~pF, $580 \pm 30 $ $e^-$ for $C=92$~pF, $770 \pm 30 $ $e^-$ for
$C=210$~pF, and $1420 \pm 30 $ $e^-$ for $C=480$~pF.  For our typical
mode of operation, the ENC should be below 1000~$e^-$, implying an
upper bound on the input capacitance at $\sim 400$~pF. Hence, the
Kapton foil anodes are hardly compatible with an area larger than a
square-meter, without a degradation of the noise performance.

\begin{figure}[h!]
    \centering
    \includegraphics[width=0.6\textwidth]{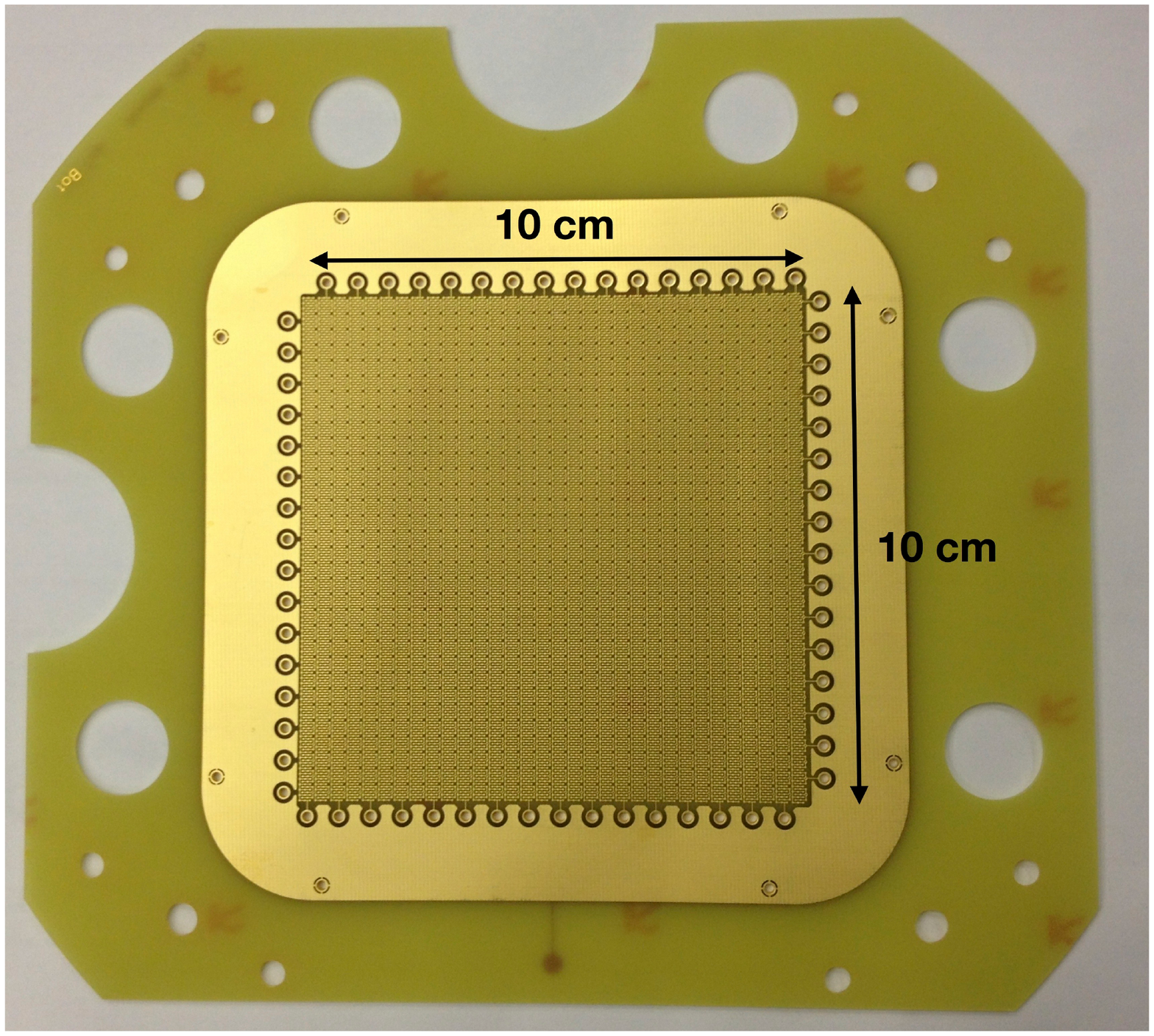}
    \caption{Picture of one of the 10x10 cm$^2$ multilayer PCB anodes
      (here anode A) tested in our chamber.}
    \label{fig:anode_overview}
  \end{figure}

\begin{figure}[h!]
    \centering
    \includegraphics[width=.8\textwidth]{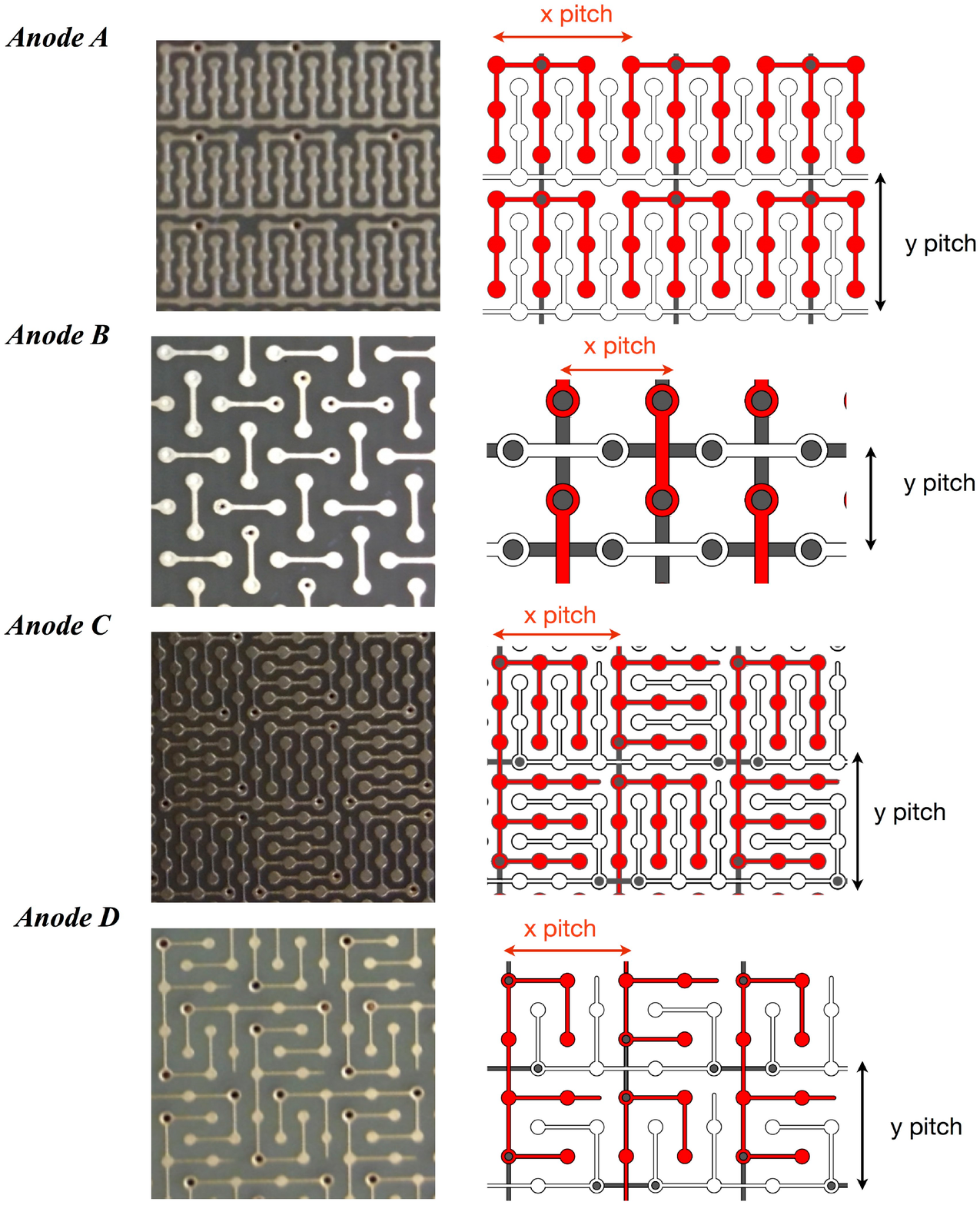}  
    \caption{Close up pictures of the 2D anode prototypes showing the
      copper track pattern that allows a 2 view readout on the same
      circuit board. Schematics explaining the interconnections
      between both views are shown on the right. The 3 mm readout
      pitches are indicated by arrows. View 0 is filled in red and
      view 1 in white.}
    \label{fig:anode_pic}
  \end{figure}

In light of those requirements, we designed new types of anodes with
both views printed on a single Printed Circuit Board (PCB) layer.
Such anodes can be easily produced using photolithographic etching
techniques (See Figure~\ref{fig:anode_overview}).  Four different
types of anodes labelled A,B, C and D (see
Figures~\ref{fig:anode_pic}) were commercially
manufactured\footnote{Anodes A, C and D were produced by Multi-CB
  (\url{www.multi-circuit-boards.eu}) and anode B by ELTOS
  (\url{www.eltos.com})}. Their dimensions and properties are listed
in \tabref{tab:anode_dim}. The 3 mm strips consist of interconnected
gold plated copper tracks. The anodes mainly differ by the number of
tracks per readout strip.

Anode A is not perfectly $x-y$ symmetric since for one view the connection between
the tracks is printed on the readout side while for the other view the
tracks are linked on the other side by a {\it via}.  Anodes B, C and D on
the other hand are designed to be fully symmetric for both
views. Since both views are linked by {\it vias}, they are printed on a
multilayer PCB. In the case of anode B, the track pitch matches the
readout pitch of 3 mm and therefore it is the anode that offers the
lowest capacitance per unit length. However with such a coarse track
pitch, the charge collected from an ionising track may not be equally
shared between both views (as will be shown later). In that respect anodes C and D were also
tested. They are similar in design to anode A but the
inter-connections between the pads are symmetric for both views. Anode
C has three copper tracks per strip and therefore has similar
capacitance per unit length as anode A. Anode D has two tracks per
strip and hence offers a lower capacitance. The measured capacitance
per unit length for each anode is reported in
\tabref{tab:anode_dim}. Compared to the Kapton foil anode,  the
electrical capacitance per channel of the new anodes is reduced by a 
factor 2 to 6.

\begin{table}[h!]
\renewcommand{\tabcolsep}{3mm}
\centering
    \begin{tabular}{lllll}
      \hline
      Parameter &  Anode A & Anode B & Anode C & Anode D\\
      \hline
      Copper thickness ($\mu$m)& 35 & 35 & 35 &35\\
      Epoxy thickness (mm)& 1.5  & 5.1  & 1.5 &1.5\\
      Readout pitch (mm)& 3  & 3.125  & 3 & 3 \\
      Track pitch (mm)& 1& 3 & 1& 1.5\\
      Track width (mm)& 0.1& 0.3 & 0.1& 0.1\\
      Pad diameter (mm)&0 .4  & 1 & 0.4 &0.4\\
      Via diameter (mm)& 0.2  & 0.3  & 0.2 & 0.2\\
      Measured capacitance  (pF/m) & 230  & 100 & 260 &140\\
      \hline
\end{tabular}
\caption{\label{tab:anode_dim}Characteristics of the 2D PCB anodes. For anodes C and D the
        capacitance was measured from a full $50\times10$ cm$^2$ PCB,
        while for anodes A and B the
        measurements are extrapolated from the $10\times10$~cm$^2$ boards.
        The copper thickness corresponds to the approximate layer after
        the final etching phase.}
\end{table}

\subsection{Single extraction grid}\label{sec:singe_grid}
\begin{figure}[t]
    \centering
    \includegraphics[width=.65\textwidth]{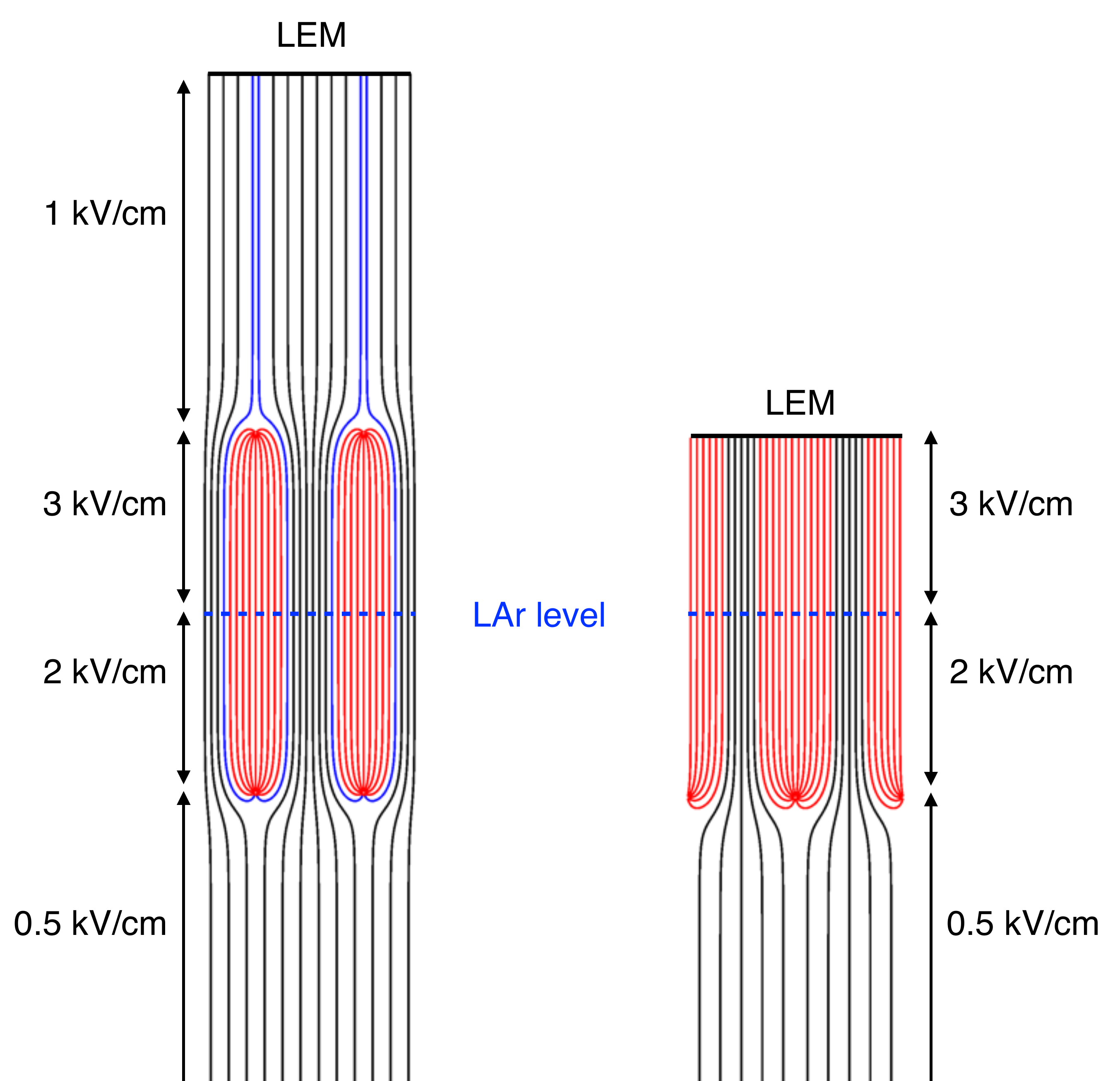}  
    \caption{Electric field lines at the liquid-gas interface for the
      double grid (left) and single grid (right) configuration. The
      field lines followed by the drifting charge are drawn in black.}
    \label{fig:grid_comsol}
  \end{figure}

Electric fields of the order of 2 kV/cm in liquid are needed to
efficiently extract the electrons from the liquid to vapor phase.  In
our previous setups \cite{Badertscher:2011sy} the extraction field
region was confined between two parallel 100 $\mu$m diameter wire
grids placed with a gap of 10~mm across the liquid-vapor
interface. Electrons may however be collected on the top grid due to
their diffusion in the gas phase, and a potential misalignment of the
two grids leads to the consequence that the grid system is not fully
transparent for the drifting electrons. To resolve these issues, we
tested extraction with a single grid placed in the liquid. In
this configuration the LEM is positioned just above the liquid surface
and the extraction field is directly provided by the LEM-grid system
over the 10~mm distance. Electrostatic calculations of the field lines
at the interface are shown in \figref{fig:grid_comsol} for a double-
and single-grid extraction. In the former, the spatial distribution of
the charge is maintained in the gas since the electrons are first
focalised between the grids and de-focalised to the original pattern
after the second grid.  In the single grid geometry, the charges stay
bunched as they arrive to the LEM and the size of the bunches is given
by the wire pitch.  Opting for a single grid system offers the extra
advantage of simplifying the overall design and construction of the
readout system.  Furthermore it reduces the anode-grid distance and lowers the
absolute value of the required high voltage.

In our test, the extraction plane was constructed with a 1.5 mm pitch
mesh, etched from a stainless steel plate. The strips had a
width of 150 $\mu$m. With this configuration and with the electric
fields indicated in \figref{fig:grid_comsol}, all electrons
are effectively transported towards the LEM. 

\section{Operation and performance assessment}
The cosmic data collected with each anode is summarised in
\tabref{tab:anode_data}. 
To study long-term stability, we decided to operate the chamber
equipped with the different anodes at an asymptotically stable gain of $G_\infty\sim 15$
(see Section~\ref{sec:lifetime} for the definition of $G_\infty$).  The
electric field applied across each stage that were set for the long term
operation are reported in
\tabref{tab:efield_config}. We could verify that under this condition,
the setup is very stable.  In a total 46~days of operation, a
comprehensive set of 14.6~million triggers has been collected and only
8~discharges observed. Further tests will be performed in the future
to assess the long-term stability at higher gains.  The results
obtained with anode A are reported in Section~\ref{sec:resa}.  The
comparison between data collected with other anode geometries is
discussed at the end of the paper in \secref{sec:uniformity}.

\begin{table}[htb]
\renewcommand{\arraystretch}{1.1}
\renewcommand{\tabcolsep}{1.2mm}
\begin{center}
\begin{tabular}[\textwidth]{rllcccccc}
\toprule 
&  \phantom{a}&\multicolumn{3}{c}{data taking period}&\phantom{a}&number
 of triggers & \phantom{a}&number of discharges\\
 \cmidrule{3-5}
& \phantom{a}& start& stop& days running& \phantom{a}&
&\phantom{a}& \\
 \midrule
 anode A& &11-Apr & 16-May & 20 & &7.5 M& &6\\
 anode B& &16-July & 29-July & 12 & &4.2 M& &2\\
 anode C& &27-Aug & 02-Sept & 7 & &1.4 M& &0\\
 anode D& &15-Oct & 21-Oct & 7 & &1.5 M& &0\\
 \bottomrule 
\end{tabular}
\end{center}
\caption{\label{tab:anode_data}Data taking periods for each anode.}
\end{table}

\begin{table}[htb]
\renewcommand{\arraystretch}{.9}
\begin{center}
\begin{tabular}[\textwidth]{lcccc}
  \toprule 
  &\phantom{ab}& distance [mm] &\phantom{ab}& nominal electric field [kV/cm]\\
\midrule
Induction (anode-LEM)& & 2 && 5 \\
Amplification (LEM)& & 1 && 33 \\
Extraction (LEM-grid)& & 10 && 2 \\
Drift field (grid-cathode)& & 210 && 0.5 \\
\bottomrule 
\end{tabular}
\end{center}
\caption{\label{tab:efield_config} Electric field configuration for the long-term
operation.}
\end{table}

\subsection{Straight track reconstruction}
Cosmic muons that cross the chamber are minimum ionising particles
   (MIPs) depositing a known amount of energy of about 2.1 MeV/cm and are 
   identifiable by events with a single ``straight track'' (plus possibly accompanying delta-rays). 
   Such events can be used to characterise the detector in terms of
   free electron lifetime and amplification. In particular the
   measurement of the electron lifetime $\tau_e$ is retrieved by
   fitting the average energy loss of the reconstructed tracks as a
   function of the drift time ($t_{drift}$) with the exponential law
   $e^{-t_{drift}/\tau_e}$.

\begin{figure}[htb]
     \centering
     \includegraphics[scale=.35]{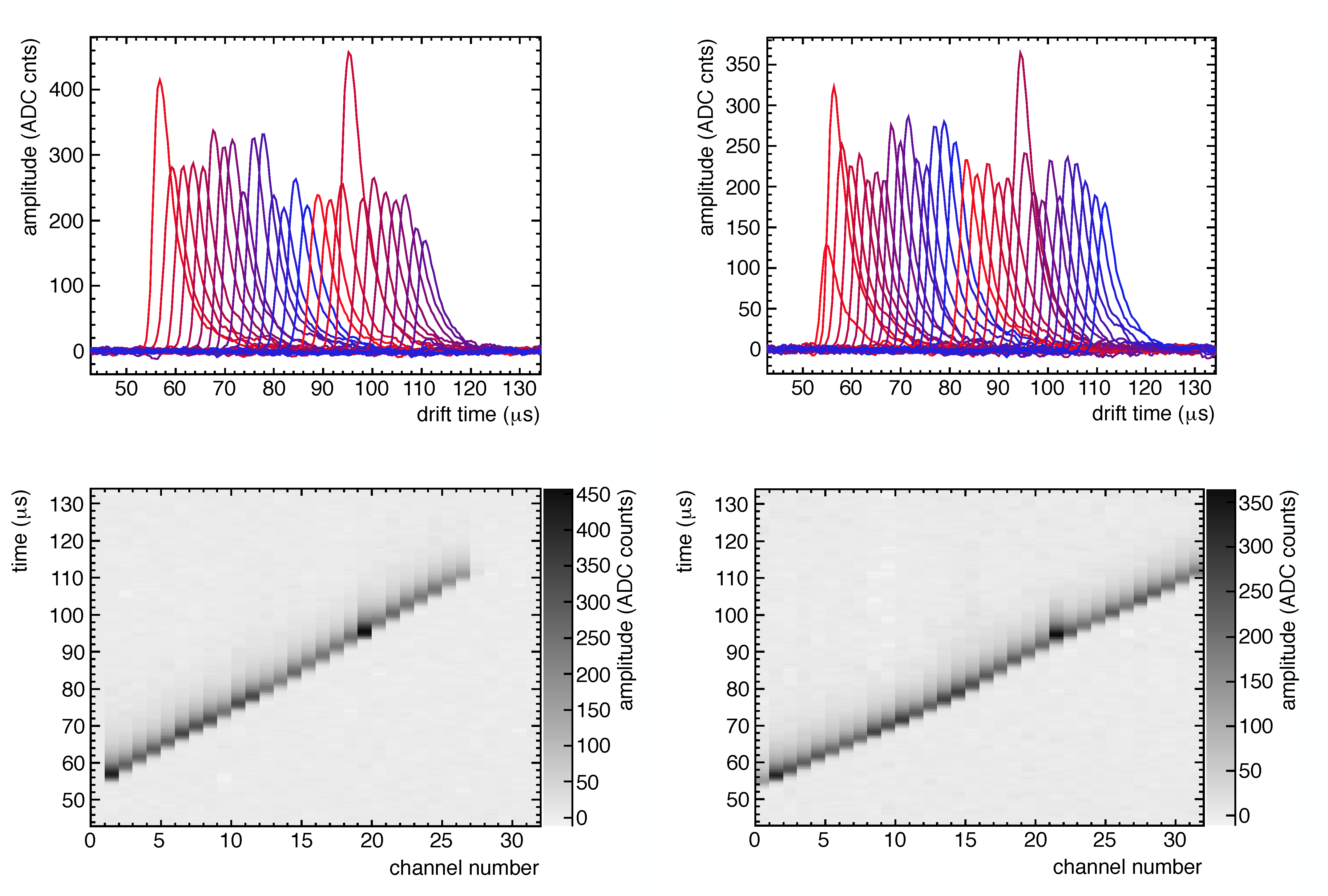}  
     \caption{Event display of a cosmic track. (top): the raw waveforms
       showing the amplitude of the signals on both views (different colours for different 
       readout channels). (bottom):
       drift time versus channel number of the reconstructed
       hits. This event was collected with anode A and the nominal
       values of the electric field settings described in the text.}
     \label{fig:cosmic_event}
   \end{figure}

\begin{figure}[htb]
     \centering
     \includegraphics[width=.7\textwidth,height=0.4\textheight]{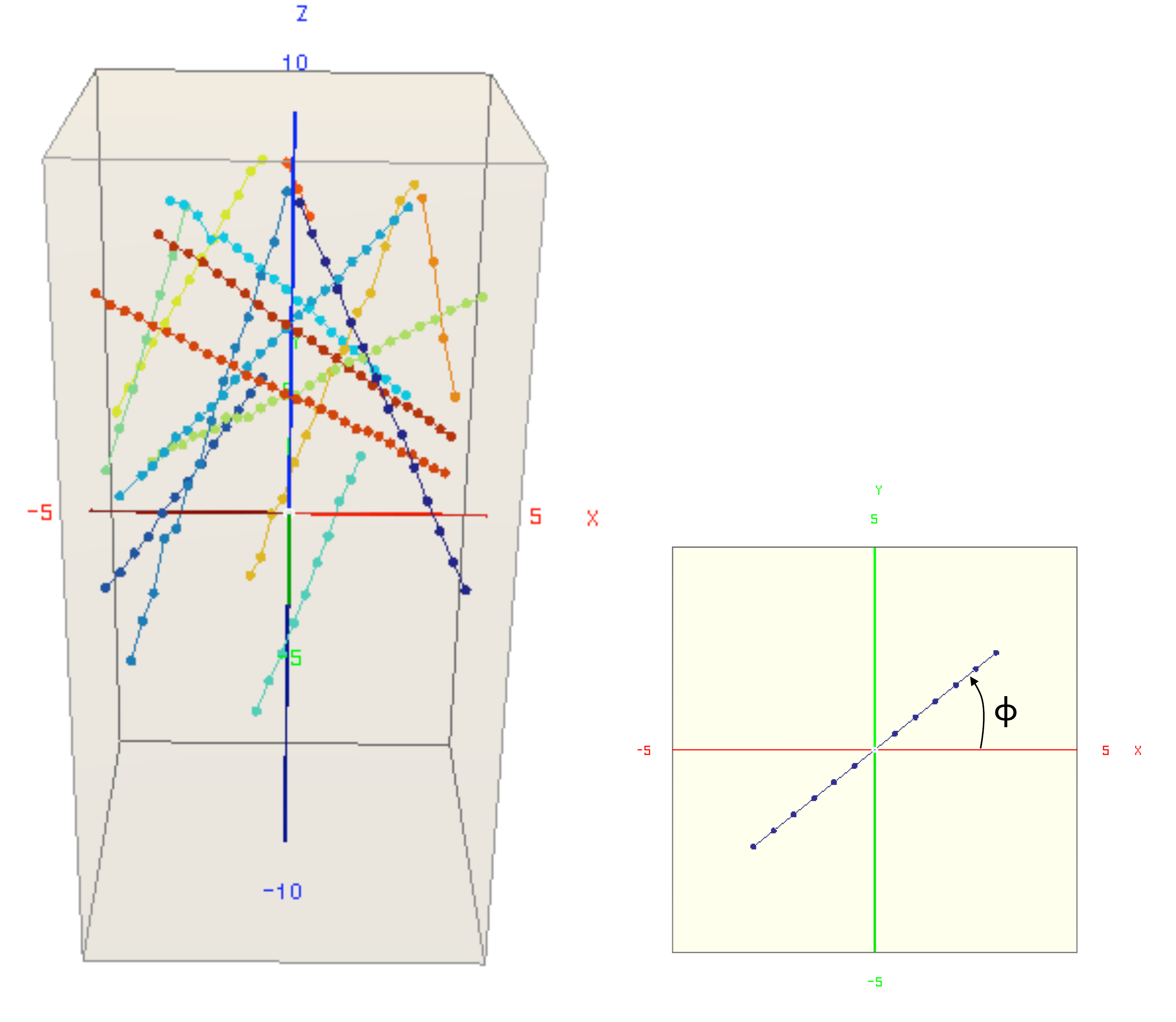}  
     \caption{Left: reconstructed 3D hits and tracks from crossing cosmic rays. The $z$
       coordinate is that of the drift and the $x-y$ coordinates are
       given by both views of the anode. Right: top view ($x-y$ plane
       ) of the detector.}
     \label{fig:3D_event}
   \end{figure}

The same offline procedure is applied to each set of data listed in Table~\ref{tab:anode_data}, and allows
for a direct comparison of the anode performance.  A typical cosmic
track event collected with the electric fields listed in Table~\ref{tab:efield_config}, 
is shown in \figref{fig:cosmic_event}. The channel number and the drift time are
plotted on the $x$ and $y$ axis respectively and the gray scale is
proportional to the signal amplitude. The 3D reconstruction of the
muon tracks from the digitised raw waveforms is accomplished with the
Qscan software package~\cite{Rico:2002}.  A set of algorithms is
consecutively applied to filter coherent pick up noise that could not
be removed online, remove discrete noise frequencies and perform
pedestal subtraction. Physical hits are subsequently extracted from
the signal waveforms by means of a standard threshold discrimination
algorithm. Once calibrated, the hit integral is a direct measurement
of the local energy deposition of the ionisation track in the liquid
Argon medium. The next step consists in identifying tracks by
clustering and fitting neighbouring hits. A tree finding
algorithm~\cite {Cassel1981235} efficiently finds tracks by searching
for the longest consecutive set of hits. The connected hits are then
fitted with a straight line and the $z$ coordinate of the 3D track is
computed by matching the end point drift times of the tracks from both
views.
   The 3D reconstruction allows to retrieve the length of the track on
   each strip of view 0 and view 1 ($\Delta s_0$ and $\Delta s_1$),
   along with the charge collected on the corresponding channels,
   $\Delta Q_0$ and $\Delta Q_1$ . The ratios \dqdxz and \dqdxo which
   are proportional to the energy locally deposited by the track in liquid
   Argon per unit length, are the relevant quantities used to estimate
   the gain of the chamber (see \equref{eq:gain}).  In addition to
   making a selection on the goodness of the linear fit, only tracks
   that cross the entire detector were retained by applying a
   selection criteria on the endpoints of the reconstructed 3D
   tracks. Examples of 3D reconstructed hits and corresponding tracks
   are displayed in \figref{fig:3D_event}.

   \subsection{Definition of the stable effective gain} \label{sec:lifetime} As
explained in Ref.~\cite{Badertscher:2011sy}, we define the effective
gain $G_{eff}$ of the device as the ratio of the measured charge (corrected for
the finite drift electron lifetime) collected on both views to the
predicted charge deposition of an ionising particle. Under the
assumption that only MIP events are present in our selected sample,
the average charge deposition along a track, predicted by the
Bethe-Bloch formula and accounting for electron-ion
recombination~\cite{Amoruso:2004dy} is $\langle \Delta Q/\Delta
s\rangle_{MIP} =10$ fC/cm.  The effective gain is hence defined as
\begin{equation}
G_{eff}=\frac{\langle \dqdxz \rangle +\langle \dqdxo\rangle}{\langle
  \Delta Q/\Delta s\rangle_{MIP}}\label{eq:gain}
\end{equation} 
where the indices correspond to view 0 and 1.  

The measured $G_{eff}$ takes into
account the charge multiplication in the LEM holes, as well as
potential charge reduction from the liquid-vapour extraction
efficiency and from the electrostatic transparency of the grid and the LEM. 
For amplification in holes across a LEM of thickness $d$ with the nominal electric field
$E_0=V/d$, it is convenient to express $G_{eff}$ 
with the function~\cite{Badertscher:2011sy}:
\begin{equation}\label{eq:gain_func}
G_{eff}(E_0,\rho,t) \equiv  
{\cal T} e^{\alpha(\rho,\kappa E_0) x}\times {\cal C}(t)
\end{equation}
where ${\cal T}$ is the transparency;
$\alpha(\rho, E)$ is the first Townsend ionisation coefficient for the electric field $E$ and density $\rho$; 
$x$ denotes the effective amplification length which can be geometrically related
to the length of the field plateau along the hole; and 
${\cal C}(t)$ represents any time variation of the gain.
Electrostatic calculations of the LEM-hole geometry give a maximum field in the hole which is lower than  the naive $V/d$, consistent
with  a value of $\kappa = 0.95$ and an effective length in the range of $0.7$~mm for a 1~mm thick LEM.
The generalised form of the first Townsend coefficient as a function of the medium density $\rho$
and the electric field $E$ can be
approximated by \cite{Aoyama:1985}:
\begin{equation}
\alpha(\rho, E)=A \rho  e^{-B \rho/E}
\end{equation}
where $A$ and $B$ are parameters depending on the gas.
A fit to the electric field dependence of the Townsend
coefficient in the range between 20 and 40~kV/cm predicted by
MAGBOLTZ~\cite{magboltz} calculations, gives $A \rho =(3160\pm 90)$ cm$^{-1}$ and $B \rho =(136.4\pm 1.0)~$kV/cm
for pure argon at 87~K and 1~bar.

For reasons
explained later in Section~\ref{sec:resa}, the long-term behavior of our setup
is consistent with an  electrostatic
charging up of the
dielectric medium of the LEM during operation, which reaches a stable condition after
a characteristic time for charging-up. After setting a particular electric field configuration,
an initial reduction of $G_{eff}$ is observed, which is reproducible and characteristic
of the setup.
Empirically we define the ``stable effective gain'' $G_\infty$ as:
\begin{equation}\label{eq:ginfinity}
G_{eff}(t)\simeq  G_\infty \times \frac{1}{(1-e^{-t/\tau})}
\end{equation}
where $\tau$ is the characteristic charging-up time, to be
determined from the data.

 
 \section{Results with anode A and stability of the gain}\label{sec:resa}\label{sec:stabgain}
To quantify the long term stability of the gain, we mounted the
chamber with a given anode and observed the evolution of the gain over
a designated period. The duration of the data taking with each anode
is listed in \tabref{tab:anode_data}. The run with anode A was the
longest and is therefore described in this section as example.

\begin{figure}[htb]
  \centering
 \includegraphics[width=.9\textwidth]{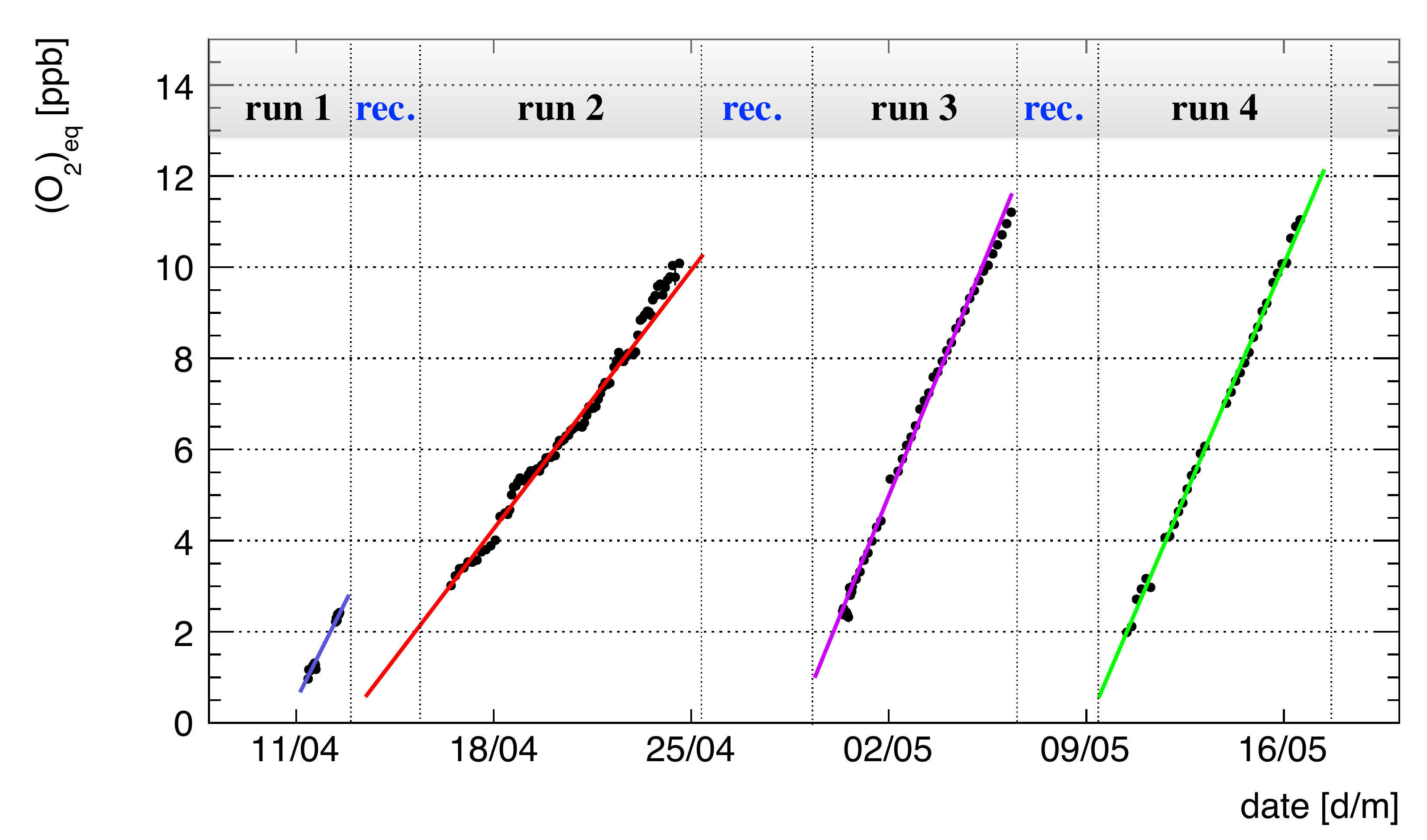}
  \caption{ Evolution of the oxygen impurity
    during the data taking period with anode A. Linear fits to the measured points
    give purity losses of about 1.4 ppb per day.}
  \label{fig:lifetime}
   \end{figure}

The data-taking with anode A lasted for more than a month at a constant event trigger rate of
about 5 Hz. During this period data was collected almost continuously,
without any changes in the settings and without
opening the chamber. The data
acquisition was interrupted three times for a few days in order to
purify the liquid Argon. The measured level of impurities in the Argon
(in Oxygen equivalent) estimated from the electron lifetime $\tau_e$,
is presented in \figref{fig:lifetime} over the full period of data-taking.  At the beginning of the
detector operation the electron lifetime was around 360 $\mu$s, which
corresponds to an oxygen equivalent impurity of $\sim$1 ppb.  During
the period of data-taking, the gas recirculation is switched off, and
a degradation of the purity of about 1.4 ppb per day is observed. The
origin of the Argon contamination is interpreted as stemming from
outgassing of the surfaces of the various components of the
chamber. Once the liquid Argon contamination reaches a level of about
10~ppb oxygen equivalent the liquid Argon is purified for a period of
3-5 days.

\begin{figure}[t]
     \centering
     \includegraphics[width=.9\textwidth]{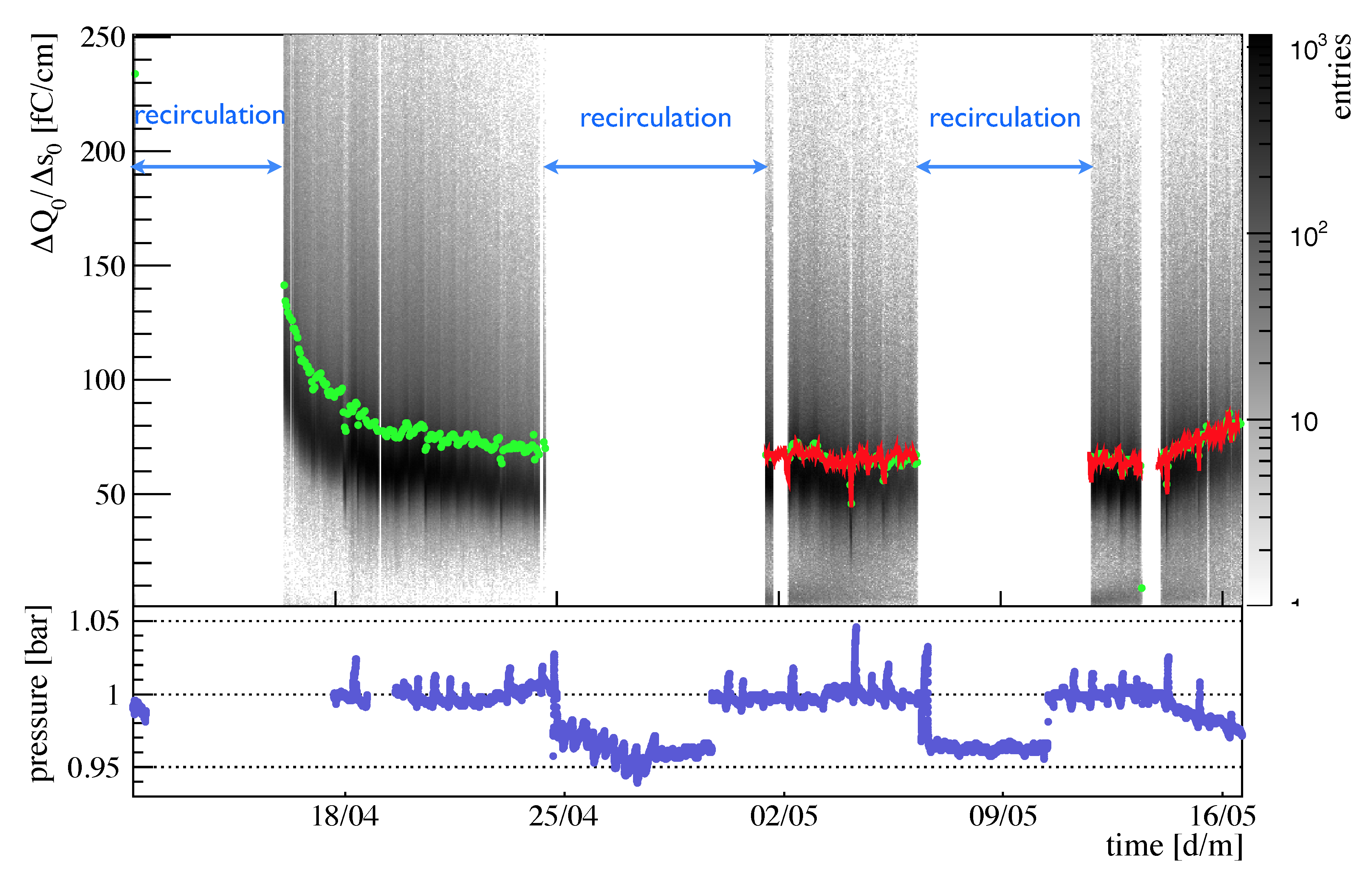}  
     \caption{Evolution of the collected charge per unit length on
       view 0 (\dqdxz) over the duration of the run. The green markers
       indicate the mean \dqdxz values and the monitored pressure of
       the gas is shown on the bottom plot. The expected fluctuations
       of the gain resulting from pressure variations are well
       described by \protect\equref{eq:gain_func}. (in red).}
     \label{fig:stabrun}
   \end{figure}

   The evolution of the measured ionisation charge per unit length
   from view 0 (\dqdxz) over the entire period is shown in
   \figref{fig:stabrun}.
The mean
\dqdxz shown by the green markers on Figure~\ref{fig:stabrun},  is proportional to the effective gain (see
\equref{eq:gain}), and is seen to stabilise after an initial decrease. 
The observed gain decrease is attributed to the electrostatic
charging up of the
dielectric medium of the LEM during operation. 
For long-term operation, the gain stabilises to the value defined as $G_\infty$.
The gain variations as a function of the gas density (i.e. related to the variations
of the pressure and the temperature
of the setup)
are well described by the estimated value of
\equref{eq:gain_func}, computed using the monitored pressure and a
constant LEM electric field, as shown by the red curve in 
Figure~\ref{fig:stabrun}.

In order to
quantitatively estimate the electrostatic charging up time of the system, the three
purification periods, during which no electric field was applied
across the LEM are removed from the graph and the time is offset to
the start of the data taking ($t_0$). 
A fit to the data points with \equref{eq:ginfinity} gives a gain which stabilises at $G_\infty\approx 15$
after an initial decrease with a characteristic time of $\tau\approx 1.6$
days.

\begin{figure}[t]
     \centering
 \includegraphics[width=.9\textwidth]{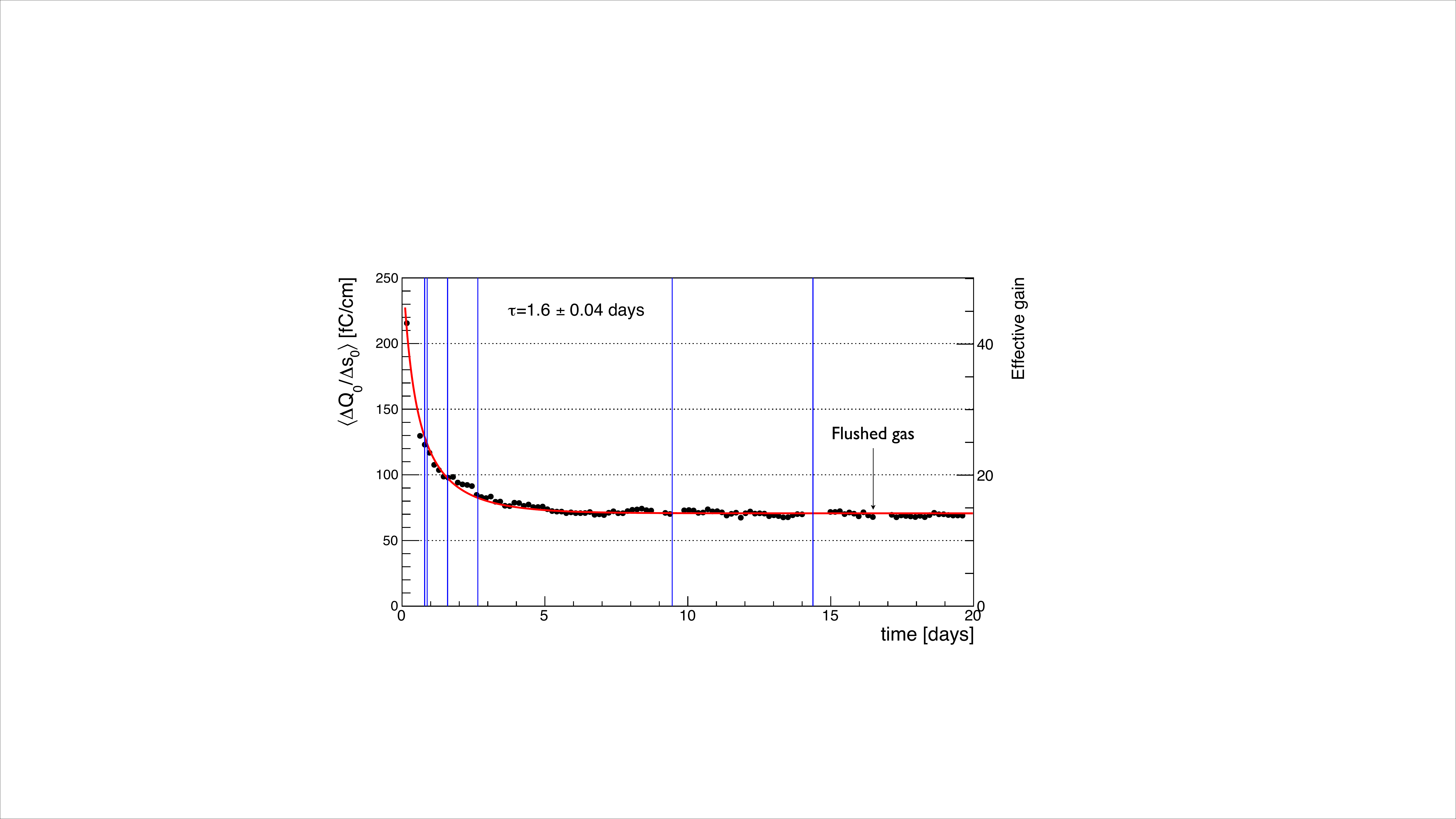}  
     \caption{Evolution of the effective gain corrected for pressure
       variations (all data have been reported with \protect\equref{eq:gain_func} to a pressure of 1~bar). 
       The data points are fitted with the function
       $G_\infty  \times \frac{1}{1-e^{-t/\tau}}$. The blue lines indicate
       the times at which discharges occurred.}
     \label{fig:gain_and_SNR}
   \end{figure}

Gain variations resulting from pressure
changes are also well fitted by the function. 
As mentioned above, the
pressure rapidly increased in our setup when the liquid Argon level in the
surrounding open bath dropped below the inner vessel top flange. It was
therefore necessary to constantly re-fill the outer bath with liquid
Argon to maintain a stable pressure inside the detector. Nevertheless the
measured charge per unit length can be corrected offline for pressure
variations which allows to compute the effective gain independently of
the pressure as shown in \figref{fig:gain_and_SNR}. 

   As indicated on \figref{fig:gain_and_SNR} by the arrow ``flushed gas''
   at one point during the
   run, the gas inside the chamber was pumped out. This procedure has
   the effect of removing nitrogen traces n the vapour phase
   and other contaminants that are not removed by the getter. Since this
   operation had no impact on the gain, we could rule out the
   possibility that the initial gain degradation was due to impurities
   slowly diffusing from the liquid into the gas phase that were not
   trapped by the purification cartridge.
   
   Once the gain stabilises, switching off the electric fields  for a limited period
   of time (e.g. during the argon purification) does 
   not promptly restore the value of the initially reached gain, and the gain
   stays at $G_\infty$. This is compatible with the hypothesis 
   of the charging up of the LEM dielectric. Likely, the charges stick on the dielectric even in  
   absence of the electric field, and argon vapour, inert and not electronegative, is not able to attach 
   charges and remove them from the surface. 

\begin{figure}[htb]
     \centering
     \includegraphics[width=.7\textwidth,scale=1]{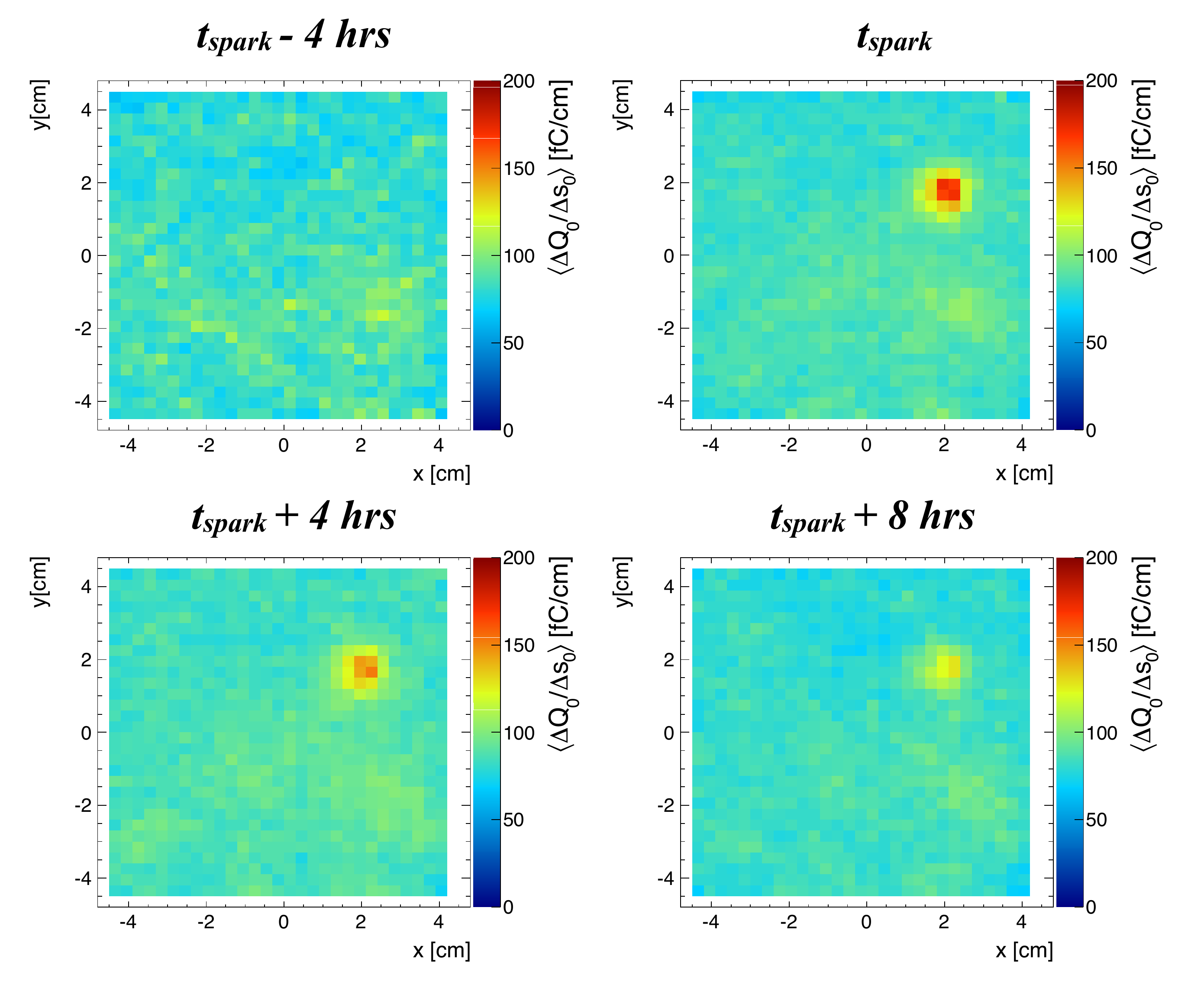}  
     \caption{Time evolution of $\langle \dqdxz \rangle$ as a function
       of the $x-y$ coordinates. The plots are computed in intervals
       of four hours from top-left to bottom-right. A discharge
       occurred immediately before the second plot.}
     \label{fig:spark_evolution_xy}
   \end{figure}

\begin{figure}[htb]
     \centering
     \includegraphics[width=.7\textwidth,scale=1]{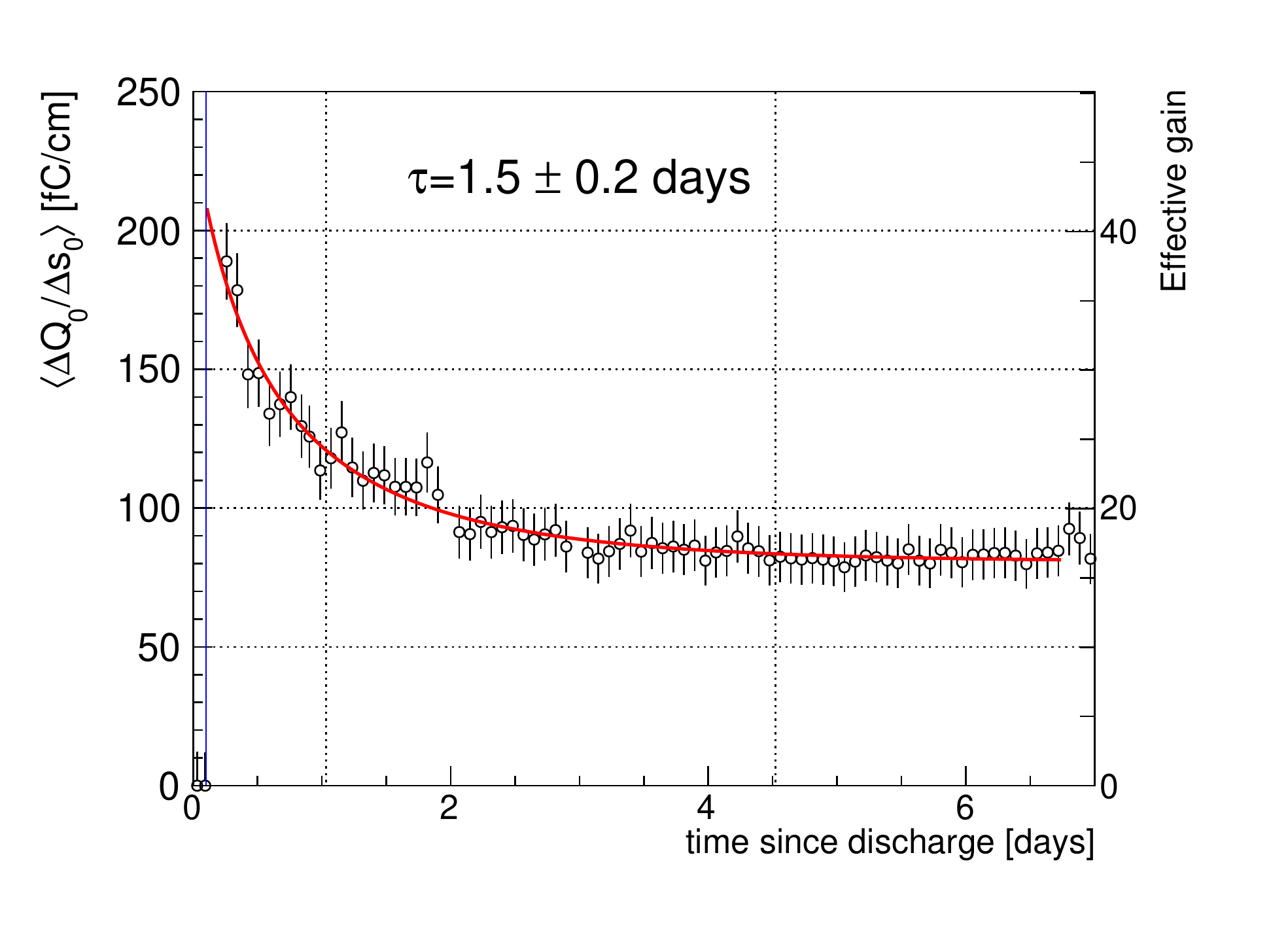}  
     \caption{Evolution of the maximum $\langle \dqdxz \rangle$ amplitude in a 1 cm$^2$
       area around the LEM hole where the discharge occurred.}
     \label{fig:spark_evolution_amp}
   \end{figure}

   During
   the  data-taking with anode A, six discharges occurred across the
   LEM, they are indicated by blue lines in
   \figref{fig:gain_and_SNR}. While it is clear on the figure that
   those discharges do not disturb the evolution of the overall gain,
   locally the gain is affected in the region where the discharge
   occurred.  This is depicted in \figref{fig:spark_evolution_xy}
   where $\langle \dqdxz \rangle$ as a function of the reconstructed $x$ and $y$
   coordinate is shown. Once a discharge occurred across a LEM hole,
   the gain at $t_0$ is recovered over a region of about 1 cm$^2$
   around the hole. As shown in \figref{fig:spark_evolution_amp} the
   locally recovered gain then decreases with a similar time constant
   of one and a half days which supports the hypothesis that the effective gain
   reduction is a consequence of the charge accumulation on the
   dielectric of the LEM. The discharge locally removes the accumulated
   charges and the area then exhibits a higher gain immediately after, until
   charges accumulation restores the stable effective gain. This behaviour
   could be reproducibly observed at each discharge of the LEM. This leads us to
   believe that the interpretation that the electrostatic charging up of the
   LEM leads to the observed decrease of the effective
   gain, is adequate.

  The signal-to-noise ratio for minimum ionising tracks is defined as the mean
   amplitude of the waveforms produced by the cosmic tracks divided by
   the average value of the noise RMS. Given our noise value of about
   4-5 adc counts RMS and with the effective gain $G_{\infty}\sim 15$,
  the chamber is in a stable operation mode with a $S/N\sim 60$ 
  for minimum ionising particles. 
  
 At the very beginning of the data taking with anode A, before the gain had
   stabilised, (i.e beginning of run 1 in \figref{fig:lifetime}) we
   briefly tested the response of chamber in terms of effective gain
   for various electric fields applied across the LEM while keeping
   the others at the values listed in \tabref{tab:efield_config}. We achieved
   a maximum effective gain of $G_{eff}\sim 90$ by ramping the amplification field
   up to 35 kV/cm. The field was increased further until the breakdown
   voltage of 36 kV/cm was reached (sparks occurred across the
   LEM). A gain of 90 corresponds to a signal-to-noise
   ratio $S/N$ of about 400 for minimum ionising particles, or
   $S/N\approx10$ for an energy deposition of 15~keV on a single
   readout channel.  
Further studies will
   be performed to check whether the chamber can be continuously
   operated at larger LEM fields and if higher gains in stable
   conditions can be reached.


   \section{Results with anodes B, C and D}\label{sec:uniformity}
   The results reported previously were from data taken with anode
   A. In this section we compare data collected from four independent
   runs each performed with one of the anodes presented in
   \secref{sec:anode_comp}. All the runs were operated at the same
   electric field configuration presented in
   \tabref{tab:efield_config}. For the chosen electric fields (same as those
   set for anode A), 
   the gain systematically exhibited the
   initial decrease reported in the previous section and stabilised at
   a value $G_\infty \sim 15$, supporting the interpretation that this behaviour
   is originating in the LEM and is independent of the anode. 
   The data used to characterise the anodes is
   corrected for pressure variations and was collected once the gain
   was stable.
   
\begin{figure}[tb]
     \centering
     \includegraphics[width=\textwidth,scale=1]{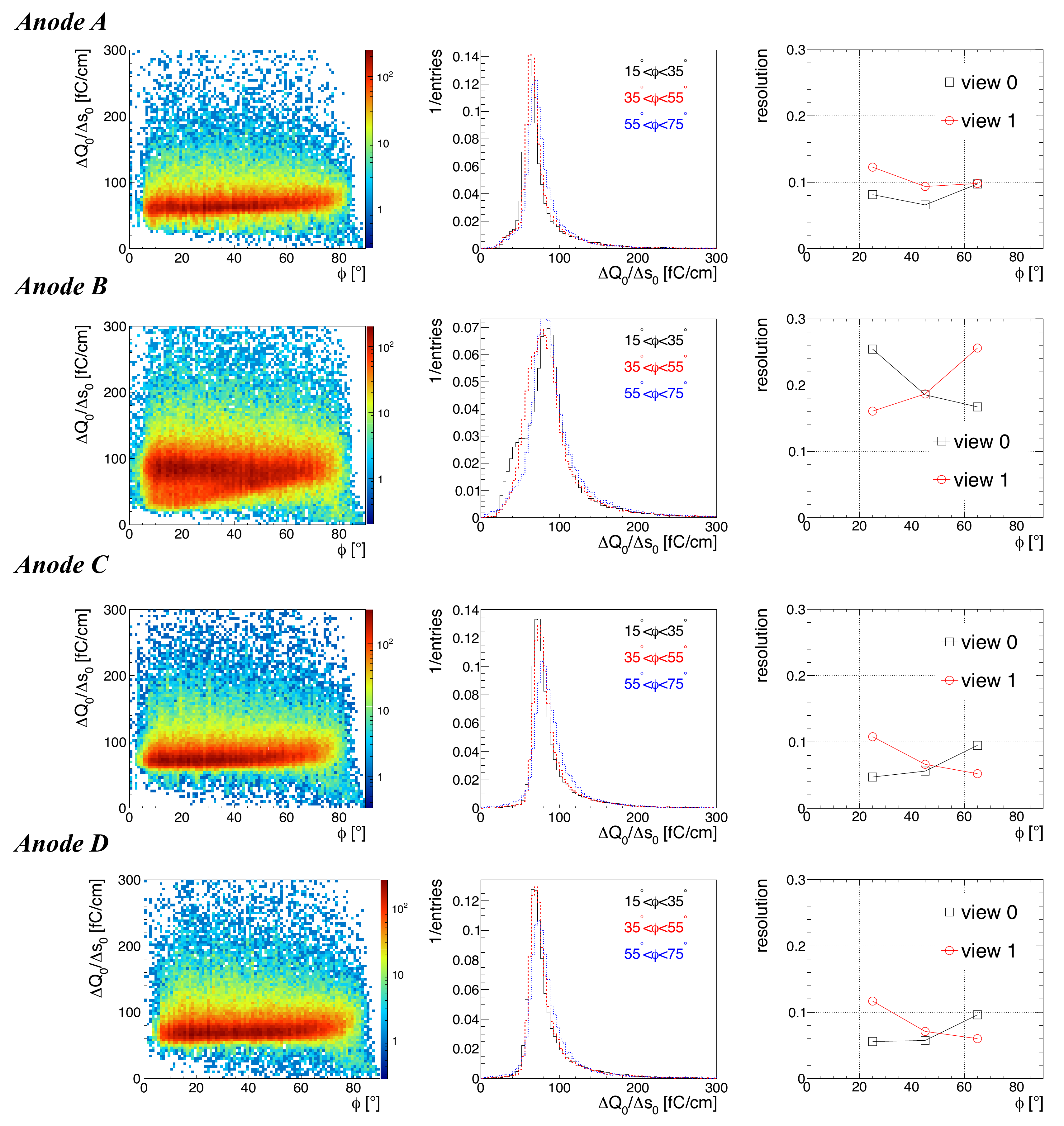}  
      \caption{\label{fig:charge_phi}Charge deposition measured on
       view 0 (\dqdxz) as a function of the track angle $\phi$ (left)
       and projection of the \dqdxz distribution in three $\phi$
       intervals (middle). The right plot shows the resolutions of
       those \dqdxz distributions as a function of $\phi$ for both
       views (see text for the definition of the resolution and of the
       angle $\phi$).}
   \end{figure}

   In \figref{fig:charge_phi} we show relevant
   distributions that illustrate the performance of each anode.
   The left plot shows the collected charge per unit length on view 0
   (\dqdxz) as a function of the track azimuthal angle $\phi$, which
   corresponds to the angle at which the tracks cross the readout
   strips (see \figref{fig:3D_event}). Distributions corresponding to
   the projections of the \dqdxz in three specified $\phi$ intervals
   are shown in the middle.  The right plot gives the resolution of
   the distributions in all three angular intervalls for both
   views. The resolution is obtained by fitting the distributions with
   a Gaussian convoluted Landau function and is defined as
   $\sigma_{gauss}/\langle\dqdxz\rangle$. 
   
\begin{figure}[htb]
     \centering
     \includegraphics[width=.8\textwidth]{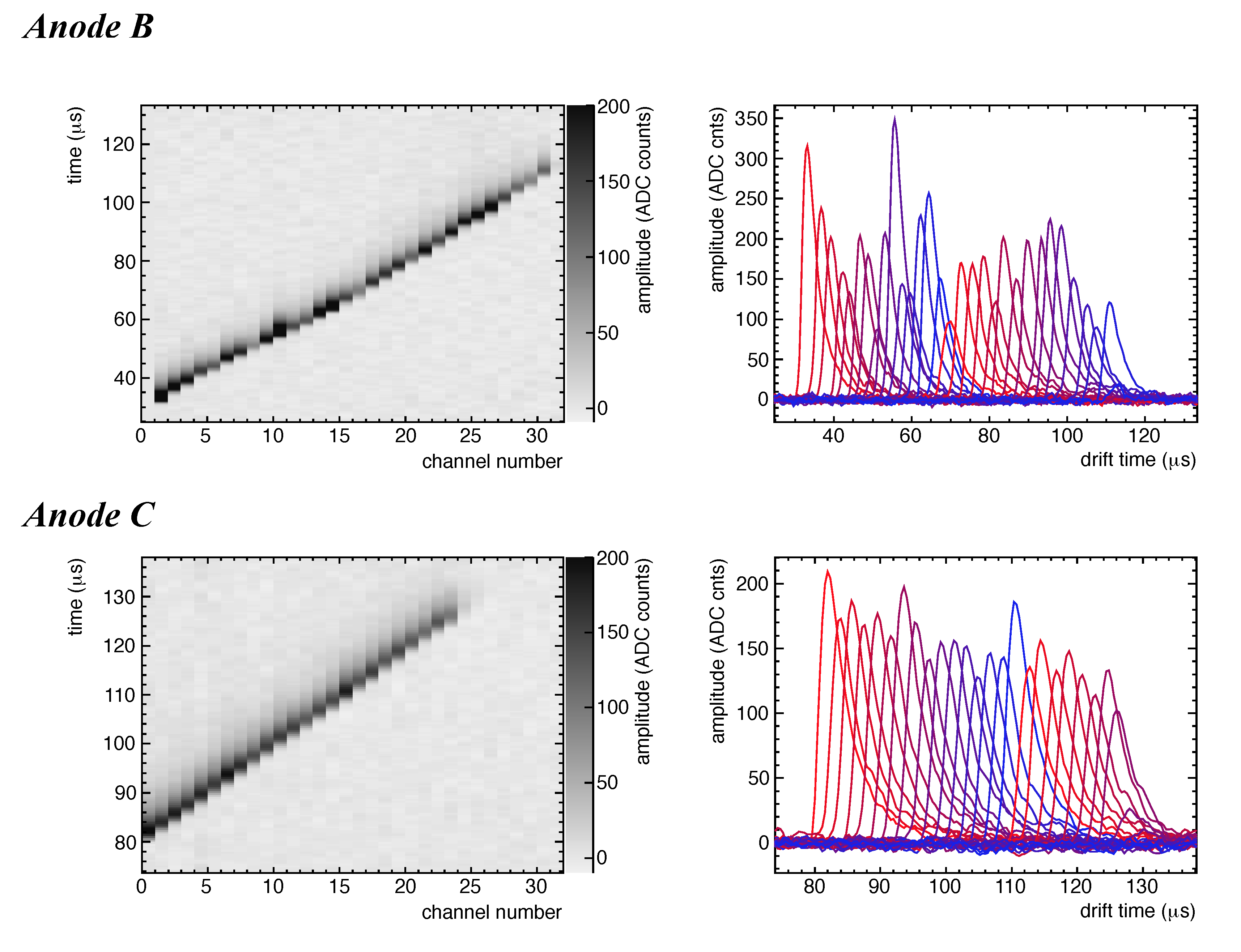}  
     \caption{Two similar cosmic muon
       events acquired with anode B and anode C (only view 0 is
       shown). Because of the coarser track pitch of anode B, the
       collected charge varies significantly between neighboring strips. As a result the
       amplitudes of the waveforms have large variations.}\label{fig:evt_display_comp}
   \end{figure}

   For all the anodes the \dqdxz distributions are centered around
   $\langle\dqdxz\rangle\approx 75$ fC/cm demonstrating the
   reproducibility of the detector operation with a stable gain around
   15. The values of the resolution between both views are in general
   symmetric around $\phi=45^{\circ}$ which is reasonable considering
   that the set of readout strips belonging to one view is rotated by
   $90^{\circ}$ with respect to the other. This also illustrates that,
   with the exception of anode A, both views are completely $x-y$
   symmetric.  For anodes A, C and D the \dqdxz distributions are
   close to a Landau function as expected from the fluctuations of the
   collected charge per unit length. The shape of those distributions
   are also similar for all angular intervals. For anode B however
   the distributions are clearly much wider and their shapes are
   dependent of the angle at which the track crosses the strips. As
   explained in \secref{sec:anode_comp} anode B has only one copper
   track per strip which means that, for a crossing muon depositing a
   constant charge per unit length, neighbouring strips may not
   collect the same amount of charge. Charge is not lost, but the sharing
   locally lacks uniformity between the two views. 
   The coarse track pitch of anode
   B therefore introduces large variations on the signals between
   neighbouring strips and consequently important fluctuations on the
   distributions of collected charge per unit length. This effect is
   illustrated in \figref{fig:evt_display_comp} where two similar
   cosmic muon events are shown, one is acquired with anode B and the
   other with anode C. Anode C has three copper tracks per strip or
   equivalently a track pitch of 1 mm. It is clear from the amplitudes
   of the waveforms that the charge collected on anode B differs
   rather strongly from strip to strip. The direct consequence of
   those large variations is a degradation on the resolution of the
   collected charge per unit length.

\section{Summary of results}
In \tabref{tab:anode_para} the properties of each anode are
summarised. For comparison those from the Kapton foil anode are also
presented in the table where the values are either taken or computed
from the data described in \cite{filippo_thesis}. Besides the other
parameters, it is important to verify that the charge is equally
shared between both views. For that matter we define the asymmetry
factor as the distribution of the difference between the charge
collected on both views normalised to their sum. The mean and RMS of
those distributions are reported in \tabref{tab:anode_para}.  
\begin{table}[h!]
\renewcommand{\arraystretch}{1.1}
\renewcommand{\tabcolsep}{1.2mm}
\begin{center}
\begin{tabular}[\textwidth]{rlcccccccc}
\toprule 
anode & &\phantom{a}& capacitance [pF/m]
 &\phantom{a}&\multicolumn{2}{l}{\dqdxi resolution [\%]}&\phantom{a}&\multicolumn{2}{l} {Asymmetry [\%]}\\
 \cmidrule{6-7}
 \cmidrule{9-10}
 &&&& & view 0& view 1& &mean&RMS\\
 \midrule
 multilayer PCB & A& & 230 && 9.3  & 11.3& &2.1&8.8\\
                        &B& & 100 && 22.9  & 22.4&&0.9& 18\\
 & C& & 260 & &5.5 & 6.4& &0.6&9.8\\
 & D& & 140 && 6.7 & 7.9& &0.7&9.0\\
 Kapton foil & & & 600 && 6.2 & 8.1 & &2.5&20\\
 \bottomrule 
\end{tabular}
\end{center}
\caption{\label{tab:anode_para}Summary of the results obtained for all
  the tested anodes. The data used for the
  Kapton foil anode is taken from \cite{filippo_thesis}.}
\end{table}
All the
anodes show good charge sharing with a mean asymmetry around the
percent level. Charge sharing is slightly worse for anode A which is
not perfectly symmetric in the $x$ and $y$ coordinates (see
\secref{sec:anode_comp}). Anode B which benefits from a low
capacitance per unit length suffers from a poor resolution on the
energy loss measurement for the reasons described previously. Final
designs of large area readouts will be motivated by using anodes with
low capacitance per unit length which still provide the best possible
resolution on the energy loss measurements. From that perspective
anode D shows the best performances.


\section{Conclusion}
\label{sec:conclusions}
We have successfully tested the characteristics of novel designs of 2D
 readout anodes manufactured from a single multilayer
printed circuit board. Since they are robust and relatively easy to
produce, they should be well suited for large area readouts. Another key
feature is that the readout strips have a typical capacitance per unit
length below 200 pF/m which means that the anode can be scaled to the
square meter level without compromising the signal-to-noise ratio. 
In our effort to further simplify the design, we have shown that the
electrons can be efficiently extracted from the liquid to the gas
phase by means of a single grid placed in the liquid. The LEM is
placed just above the liquid level and the extraction field is
provided by the LEM-grid system. 

A chamber newly equipped with the
printed circuit board anode and the single extraction grid has
achieved a high level of performance: in order to study the long-term stability,
we operated the detector for a total duration of 46~days.
We reproducibly observe that after an initial decrease
  with a characteristic time of $\tau\approx 1.6$ days, the observed
  gain is stable. 
A stable effective gain (corrected for pressure variations) of $\sim 15$ was observed
and 14.6~millions trigger collected. 
For the first time, a maximum gain of 90 was reached with an impressive $S/N\simeq 400$ for minimum ionising
particles. 

Future readout planes for the GLACIER design, as proposed for instance
for LBNO~\cite{Stahl:2012exa}, would consist of individual square meter modules
encompassing the extraction grid, the LEM and the low capacitance
anode. The good quality of the data presented in this paper is
therefore encouraging for future plans to assemble and test such
square meter modules.

\end{document}